\documentclass[
amsmath,%
aps,%
pra,%
groupedaddress,
final,%
twocolumn,%
showpacs,%
nofootinbib,%
]%
{revtex4}

  \usepackage{graphicx}
  \usepackage{hyperref}

\newcommand{\rf}[1]{(\ref{#1})}
\newcommand{\rff}[1]{FIG \ref{#1}}

\newcommand{\bc}{\begin{center}}
\newcommand{\ec}{\end{center}}
\newcommand{\be}{\begin{eqnarray}}
\newcommand{\ee}{\end{eqnarray}}

\newcommand{\bfr}{\begin{flushright}}
\newcommand{\efr}{\end{flushright}}
\newcommand{\bfl}{\begin{flushleft}}
\newcommand{\efl}{\end{flushleft}}
\newcommand{\dsp}{\displaystyle}

\newcommand{\ra}{{\rho^{a}}}
\newcommand{\rai}[1]{{\rho_{#1}^{a}}}

\newcommand{\rft}{\rho_f}
\newcommand{\rftt}{\rho_f^{(1)}}
\newcommand{\rftw}{\rho_f^{(2)}}

\newcommand{\Ha}{H_{a}}
\newcommand{\Har}{H_{a}}
\newcommand{\Hfr}{H_{f}}
\newcommand{\Hf}{H_{f}^{r}}
\newcommand{\Vaf}{V_{af}}

\newcommand{\Vafo}{\tilde{V}_{af}}

\newcommand{\Va}{V_{a}}
\newcommand{\Vp}{V_{a}^{p}}

\newcommand{\Pf}{P_{f}}

\newcommand{\Spa}{^{a}}
\newcommand{\Spf}{^{f}}
\newcommand{\Spaa}{^{a^{''}}}
\newcommand{\Spff}{^{f^{''}}}

\newcommand{\gk}{g_l({\bf k}_\alpha)}
\newcommand{\gkm}{g_m({\bf k}_\alpha)}
\newcommand{\bk}{{\bf k}}
\newcommand{\bkl}{{\bf k}_\alpha}
\newcommand{\gkl}{g_l(\bkl)}

\newcommand{\rk}{{\bf r}}

\renewcommand{\ap}{\hat{a}_{\textbf{k}_\alpha}^{+}}
\newcommand{\am}{\hat{a}_{\textbf{k}_\alpha}}

\newcommand{\Dk}[1]{D(#1)}
\newcommand{\Dks}[1]{D^{+}(#1)}

\newcommand{\pto}{|2 \rangle\langle 1|}
\newcommand{\ptt}{|2 \rangle\langle 2|}
\newcommand{\pot}{|1 \rangle\langle 2|}

\newcommand{\ass}{a^{''}}

\newcommand{\fss}{f^{''}}

\begin{document}
\title{Local field effects and multimode stimulated light scattering in a two-level medium}

\author{Vl.~K. Roerich
} \affiliation{FSUE "SRC RF TRINITI", Federal State unitary
Enterprise "State Research Center of Russian Federation Troitsk
Institute of Innovation and Fusion Researches", TRINITI, Troitsk,
Moscow region, Russia 142190} \email[Corresponding author e-mail:
]{vroerich@triniti.ru}
\author{A.~A. Panteleev} \affiliation{FSUE "SRC RF TRINITI", Federal
State unitary Enterprise "State Research Center of Russian
Federation Troitsk Institute of Innovation and Fusion Researches",
TRINITI, Troitsk, Moscow region, Russia 142190}
\email[Corresponding author e-mail: ]{vroerich@triniti.ru}
\author{M.~G. Gladush} \affiliation{FSUE "SRC RF TRINITI", Federal
State unitary Enterprise "State Research Center of Russian
Federation Troitsk Institute of Innovation and Fusion Researches",
TRINITI, Troitsk, Moscow region, Russia 142190}
\email[Corresponding author e-mail: ]{mglad@triniti.ru}

\begin{abstract}
Scattering of a short (much shorter then the spontaneous lifetime)
laser pulse has been considered in a dense resonant medium subject
to local field effects. The system was studied in the limit of
Hartree-Fock approximation for Bogolubov-Born-Green-Kirkwood-Yvon
(BBGKY) hierarchy of equations for reduced density operators. A
closed set of equations for atomic and field density operators was
derived to describe stimulated scattering of radiation.
Numerically as well as theoretically the capability of
multicomponent spectrum with maximums multiple to Rabi frequency
has been demonstrated. Relative line intensities in the spectrum
were found in the limit of low density.
\end{abstract}

 \maketitle

\section{ Introduction. \label{Sec:1}}
The properties of scattered light which one can observe under
action of a strong laser field have been the area of intensive
research in quantum optics. Wigner-Weisskopf \cite{1} were the
first to demonstrate that when a two-level system is excited with
a weak monochromatic field its fluorescence spectrum was
determined by Rayleigh scattering. In this case the criterion for
the weak pump field was the ratio of Rabi frequency to either the
spontaneous decay rate or detuning from resonance. In strong
fields the stimulated (Rayleigh) component is accompanied by
spontaneous radiation displaying two additional maximums in the
spectrum building the well known Mollow triplet \cite{Mollow69}.
It must be noted that for stationary pump modes in the strong
field limit the coherent component of radiation is reciprocally
proportional to the square of Rabi frequency and is, therefore,
negligible. These result have found numerous excellent
experimental verifications \cite{Exp}.

The general picture is much more complicated when such a process
is studied in dense media. This is mainly due to strong
interatomic interactions giving rise to collective behavior of
atoms exposed to external laser fields. The pioneer consideration
of this issue was presented in work \cite{Dike} by Dicke who
demonstrated that in an optically dense medium of $N$ atoms
($n\lambdabar^3\gg1$, where $n$ is the density of atoms and
$\lambdabar=\lambda/2\pi$ $\lambda$ is the wavelength) collective
spontaneous decay with its intensity proportional to $N^2$ was
possible.

One of the most studied range of phenomena in this aspect is
linked to the interaction of closely positioned atomic systems
through their radiation field and combines a number of well
investigated cooperative effects \cite{Andreev}. This type of
phenomena includes local-field effects or the near dipole-dipole
interaction, a striking sequence of which is intrinsic optical
bistability predicted and studied theoretically \cite{Bowden_79,
Bowden_84-93, Bowden_96, Bowden_Rev} and observed in experiments
with impurities embedded into crystalline matrices or doped
glasses \cite{Crystaline, Glass}. It has been demonstrated that
local field effects can notably affect the spontaneous decay rate
\cite{Bowden_2000, Fleischhauer, Knoll}. Also, for these case the
strict condition forbidding realization of such bistability
mechanism cased by collisional broadening was eliminated
\cite{Manassah_89, Manassah_2001}.

Another area of studies is linked to collective spontaneous decay.
In addition to higher decay rate \cite{Dike} interatomic
interaction can substantially modify the spontaneous spectrum in
comparison with the single atom case \cite{Agarwal77,Amin78}.
Interaction between the atoms gives rise to additional resonances
in absorption and emission spectra which is explained by
possibility for simultaneous excitation of different atoms and
excitation exchange
\cite{Senitzky78,Carmichael79,Agarwal80,Drummond80,Freedhoff80,Ficek81,Kilin80,Kus81,Griffin82,Ficek84}.

Additional lines including satellites multiple to Rabi frequency
in superradiant spectra was first reported and discussed in
\cite{Agarwal77,Amin78,Senitzky78}. However, intensities of such
additional sidebands were shown to be utterly small, i.e., $\sim
10^{-6}$ to the central peak. It must be noted that there are some
other difficulties found in connection to  conditions for the
superradiant mode. This fact puts the limitations on the value of
the transverse relaxation constant \cite{Mandel} $\gamma_2\ll
N\gamma$, where $\gamma$ is the radiative relaxation rate. It all
makes it very complicated to observe the sidebands experimentally.

One should note that in the references cited above there was
steady state spectra analysis only so the contribution of
stimulated scattering (Rayleigh component) in resonance
fluorescence has not been considered. On one hand it could have
been conditioned by its smallness, but on the other hand in the
steady state it is scattered "elastically". However, as we
demonstrated in \cite{Roerich2003} in transient spectra the
profile of stimulated scattering is different from delta function
and its frequency is not the same as that of the pump field.
Besides, its intensity is no small in comparison with that of
spontaneous radiation. In dense media, similar to superradiation,
intensity is proportional to the square of atomic density and
under certain conditions may dominate over spontaneous intensity
especially when the superradiant mode is not likely to arise
because of collisions.

The aim of this work is to study transient spectra of stimulated
scattering in a dense medium excited by a short laser pulse. In
the limit of Hartree-Fock approximation applied for
BBGKY-hierarchy of equations for the reduced density operators
\cite{Bonitz} we will receive equations of motion for atomic and
field density operators which imply local field corrections giving
rise to nonlinearities of the system. Expressions describing
intensity spectra of scattered light will be also obtained. Using
the method of successive approximations we will conduct
theoretical analysis of scattered light in case when the medium is
excited by a short laser pulse of constant intensity profile. In
conclusion we will make a comparison of the analytical solution
with the exact solution obtained numerically and study the
spectral properties in a wide range of parameters.

This work is structured as follows. In Sec. \ref{Sec:2} we present
the approximations to be used and derive the equations of motion
for the atomic and field density operators. In Sec. \ref{Sec:3} we
apply the method of successive approximations to obtain analytical
solutions for the equations and discuss their properties. In Sec.
\ref{Sec:4} the spectrum of scattered light is studied
numerically. A comparison with the analytic solution is carried
out while the properties of obtained spectra are studied for a
wider range of parameters. In the summary we list the basic
results. The appendix contains the explicit derivation scheme for
the local field operator.

\section{Basic equations \label{Sec:2}}
We will consider interaction of a medium of nondegenerate
two-level atoms with a laser field resonant to $1\to 2$ atomic
transition. In the dipole and rotative wave approximations the
Hamiltonian of the system, describing the interaction of $N$ atoms
and laser field, in units $1/\hbar$ has the form:
  \be
    \dsp{H=\sum^{N}_{l=1}(\Ha+\Va)+\sum_{\bk}\Hfr+\sum^{N}_{l=1}\sum_{\bk}\Vaf,} \\
  \ee
where
  \be
   \begin{array}{c}
   \dsp{\Har=\Delta_{21}\ptt, \;\;\;\;}
   \dsp{\Hfr=\nu_\bk \ap\am,} \\
   \dsp{\Vaf=i(\gkl \pto \am -\gkl^{*} \pot\ap),} \\
   \dsp{\Va=i(R(t,\rk)\pto - R^{*}(t,\rk)\pot).}
  \end{array}
  \label{Gal}
 \ee
In \rf{Gal} $\Har$ describes the Hamiltonian of an unperturbed
atomic system, $\Delta_{21}=\omega_{21}-\omega_L$ is the detuning
from resonance, $\omega_{21}$ being the transition frequency and
$\omega_L$ the carrier frequency of a laser pulse,
$|j\rangle,\langle j|, j=1,2$ are the projection operators for
respective atomic states.

The next term $\Hfr$ describes the Hamiltonian of quantized
radiation field, where $\ap,\am$ are formation and annihilation
operators respectively; $\nu_\bk=\omega_\bk-\omega_L$,
$\omega_\bk$ is the frequency of a photon with the wave vector
$\bk$ and polarization $\alpha$.

Operator $\Vaf$ describes interaction of atom $l$ with a mode of
the quantized field, where  $\dsp{\gk=\sqrt{\frac{2\pi\omega_\bk}{\hbar W}}}e^{i\bk \rk_{l}}(\vec{\mu}_{21}
  \epsilon(\bkl) )$ defines the coupling constant, here $\vec{\mu}_{21}$ is the dipole transition operator,
 $\epsilon(\bkl)$ is a unitary polarization vector, and
 $\bk\cdot\epsilon(\bkl)=0$, $\alpha=1,2$
  $W$ is the quantization volume. Operator $\Va$ describes interaction of an atom
with the laser field $R(t,\rk)=-\dsp{\frac{\vec{\mu}_{21}{\bf
E}_L(t,\rk)}{2\hbar}}$, where $E_L(t,\rk)$ is the laser field
strength.

Besides, it is necessary to introduce an operator which describes
radiative absorption by the cavity walls related to either
operation of a detector device or whatever else irretrievable loss
of radiation. Allowing for the losses brings an additional term to
the Hamiltonian \rf{Gal} having the form $\sum_\bk K_\bk[\rho] $
and describing interaction of the quantized field with thermostat
at zero temperature \cite{Perelomov}:
 \be
   \label{loss}
    K_\bk[\rho]=-i\eta_\bk/2(\ap\am\rho-2\am\rho\ap +\rho\ap\am),
 \ee
where $\eta_\bk$~determines the mode losses.

In order to obtain equations describing the spectrum of scattered
light we will use the BBGKY-hierarchy in the framework of
Hartree-Fock approximation applying the well-known cluster
expansion for reduced density operators \cite{Bonitz}. While
carrying out the derivation steps we will neglect the spontaneous
radiation and collisional relaxation considering the laser pulse
duration sufficiently small. However, we take into account the
fact that the resonant medium can be dense enough, i.e.,
$n(\rk)\lambda_{21}^3/8\pi^3 \gg 1$ to let the local (cooperative)
field effects to be strong. Here, $n(\rk)$ is the density and
$\lambda_{21}$ is the transition wavelength. We will further
neglect spontaneous scattering as its intensity is low if compared
with that of the simulated scattering being proportional to the
squared density \cite{Roerich2003}.

We assume that initially the atoms are not perturbed which makes it
possible to represent the atom-field density operator as the product of atomic and field parts:
 \be
    \rho=\prod_N
    \rai{0}(\rai{11}=1)\prod_{\bkl}|0_{\bkl}\rangle\langle0_{\bkl}|.
 \ee
In the view of approximation taken above the BBGKY-hierarchy takes the
form \footnote{Unless otherwise stated the initial conditions for the systems considered remain unchanged.}:
  \be
   \label{main1}
    \begin{array}{c}
     \dsp{i\frac{d\ra}{dt}}-[\Ha+\Va,\ra]-[V_{a\fss},\ra\rho_{\fss}]\Spff=0, \\
     \dsp{i\frac{d\rft}{dt}}-[\Hf,\rft]-K_\bk[\rft]-[V_{\ass f},\rho_{\ass}\rft]\Spaa=0,\\
     \ra(0)=\rai{0},\;\;\;\;\rft(0)=|0_{\bkl}\rangle\langle0_{\bkl}|, \\
  \end{array}
\ee where $\ra,\rft$ are atomic and field single particle density
operators respectively; $[.]\Spf,[.]\Spa$ denote tracing operation
over field and atomic variables with $[.]$ being usual commutator
brackets. The BBGKY-hierarchy in this approximation is comprised
of a Bloch equation for two-level atoms interacting with laser
field complemented with the term $[\Vaf,\ra\rft]\Spf$ which is
responsible for the feedback to the atom from the scattered light
and the field equation written in Shrodinger representation. Here,
the term $[\Vaf,\ra\rft ]\Spa$ is the quantum representation of
field induced atomic polarization.

In order to find a solution for the system \rf{main1} we will use
the transformation which already used in papers
\cite{Roerich2003}. At first, however, it is expedient to change
to the wave picture by means of the substitution $\rft=e^{-i\Hf
t}\rftt e^{i\Hf t}$. As the result, the system \rf{main1} now
reads
 \be
 \label{main2}
  \begin{array}{c}
   \dsp{i\frac{d\ra}{dt}}-[\Ha+\Va,\ra]-[\Vafo,\ra\rftt]\Spf=0, \\
   \dsp{i\frac{d\rftt}{dt}}-K_\bk[\rftt]-[\Vafo,\ra\rftt ]\Spa=0, \\
   \dsp{  \Vafo=i\gk\pto \am e^{-i\nu_k t} + H.c.}. \\
  \end{array}
\ee
Let us write out explicitly the operator describing the induced atomic polarization:
 \be
 \label{oper}
  \begin{array}{c}
   \dsp{[\Vafo,\ra\rftt]\Spa=[\Pf,\rftt]=} \\
     \dsp{\sum_m^N i[\gkm\rai{12} \am e^{-i\nu_k t} + H.c.,\rftt].}
  \end{array}
\ee It follows from \rf{oper} for operator $\Pf$ that the exponent
$\exp(-i\Pf)$ is nothing but a coherent state. Allowing for the
mode losses via $K_\bk[\rftt]$, the operator which eliminates the
terms linear in $\ap,\am$ from the field equation is expressed as
follows \cite{Perelomov}:
 \be
 \label{oper2}
  \begin{array}{c}
   \exp(-i L_f)(t)=\exp(-i\phi(t))D(\beta_\bk(t)), \\
   \dsp{ \frac{d\beta_\bk}{dt}=-\frac{\eta_\bk}{2}\beta_\bk -\sum_m^N\gkm^{*}\rai
   {21}e^{i\nu_\bk t}=\zeta_\bk(t)},\\
   \dsp{ \frac{d\phi_\bk}{dt}=-i(\zeta_\bk\beta_\bk^{*}-\zeta_\bk^{*}\beta_\bk)}/2. \\
  \end{array}
 \ee
Let us note that the following ordinary operator relations for
coherent state operators are valid $D(\beta_\bk(t))$
\cite{Mandel,Perelomov}:
 \be
 \label{Zamena}
  \begin{array}{c}
   \dsp{  \Dks{\beta_\bk(t)}\am\Dk{\beta_\bk(t)}}
   \dsp{  =(\am+\beta_\bk(t)) },\\
   \dsp{  \Dks{\beta_\bk(t)}\ap\Dk{\beta_\bk(t)}}
   \dsp{  =(\ap+\beta_\bk^{*}(t)) }.\\
  \end{array}
\ee Using \rf{Zamena} while performing a transformation so that
$\rftt=\exp(-i L_f)(t)\rftw \exp(i L_f)(t)$ the system
(\ref{main2}) comes to the form
 \be
 \label{main3}
  \begin{array}{c}
   \dsp{i\frac{d\ra}{dt}}-[\Ha+\Va,\ra]-[\Vafo+\tilde{V}_{af}^{p},\ra\rftw]\Spf=0,\\
   \dsp{i\frac{d\rftw}{dt}}-K_\bk[\rftw]=0, \\
   \dsp{  \tilde{V}_{af}^{p}=i\gk(\pto \beta_\bk(t) e^{-i\nu_k t} + H.c.}.\\
  \end{array}
 \ee
Summation over atomic and field variables (see Appendix
{\bf A}) in $[\tilde{V}_{af}^{p},\ra\rftw]\Spf$
yields:
 \be
 \label{Zamena1}
  \begin{array}{c}
  \dsp{[\tilde{V}_{af}^{p},\ra\rftw]\Spf=[\Vp,\ra] },\\
   \Vp=\dsp{-i\frac{3n(\rk)G\lambda_{21}^3}{8\pi^2} \gamma/2\;\pto \rai{21}+ H.c.}.\\
  \end{array}
\ee where $\gamma$ is the radiative relaxation constant (see
Appendix); and $G$ is the geometrical factor being a function of
selected geometry of the sample, its dimensions, and polarization
of scattered light \cite{Bowden}. In its structure and nature the
operator \rf{Zamena1} is analogous to the Lorentz field or the
local field correction operator obtained in works \cite{Bowden} or
using other methods and approaches.

It is obvious that with the local field term and taking account
for the fact that for the chosen initial conditions
$K_\bk[\rftw]\equiv 0$ one can write $\rftw\equiv
|0_\bk\rangle\langle
 0_\bk|$ è $[\Vafo,\ra\rftw]\Spf\equiv 0$. Thus, the system \rf{main4} can, indeed,
be reduced to a single equation, which is
 \be
 \label{main4}
  \begin{array}{c}
   \dsp{i\frac{d\ra}{dt}}-[\Ha+\Va+\Vp,\ra]=0, \\
  \end{array}
 \ee
Changing back to the picture we started with on our way from the equations
\rf{main1} to \rf{main4} and using the properties of the coherent state operators,
the field density operator for the scattered field one can represent in the following form:
 $$
   \begin{array}{c}
  \dsp{ \rft=( e^{-i\nu_\bk\ap\am t} \Dk{\beta_\bk(t)}|0_\bk\rangle\langle 0_\bk|
   \Dks{\beta_\bk(t)}e^{i\nu_\bk\ap\am t} }),\\
    \dsp{ \rft=(\Dk{\beta_\bk(t)e^{-i\nu_\bk t}}|0_\bk\rangle\langle 0_\bk| \Dks{\beta_\bk(t)e^{-i\nu_\bk t}}
         }).\\
   \end{array}
 $$
Now, substituting $\beta_\bk(t)e^{-i\nu_\bk
 t}=\beta_\bk^{'}(t)$ we can finally write the system of equations describing the scatered light:
 \be
  \label{Res3}
  \begin{array}{c}
    \dsp{i\frac{d\ra}{dt}}-[\Ha+\Va+\Vp,\ra]=0, \\
   \dsp{  \frac{d\beta_\bk^{'}}{dt}=-(i\nu_\bk+\eta_\bk/2)\beta_\bk^{'}-\sum_m^N\gkm^{*}\rai{21}},\\
   \dsp{ \rft=|\beta_\bk^{'}(t)\rangle
   \langle \beta_\bk^{'}(t)|  }.\\
  \end{array}
 \ee
Furthermore, one can easily find from \rf{Res3} the average value of observed spectral intensity of scattered light:
 \be
   \label{Res5}
   \begin{array}{c}
   \dsp{ {I}_\bk(t) }
   \dsp{ =\hbar\omega_\bk[-iK_\bk[\rft]\ap\am]^n }=\\
   \dsp{ =\hbar\omega_\bk \langle\beta_\bk^{'}(t)|-iK_\bk[\rft]\ap\am
   |\beta_\bk^{'}(t)\rangle=}\\
   \dsp{=\eta_\bk\hbar\omega_\bk |\beta_\bk^{'}(t)|^2 }.
 \end{array}
  \ee
where $[.]^n$ is the trace over number of photons.

\section{Scattered light spectrum of a dot sample excited by a short laser pulse \label{Sec:3}}
In this section we will study the properties of the scattered
light spectrum for a dot sample excited by a short laser pulse.
The main emphasis in our analysis falls on the effects induced by
the local field correction. We will assume that the collection of
atoms is embedded in a sample with dimensions small enough to let
us drop all spatial dependencies in equations \rf{Res3}. Besides,
for the sake of simplicity we consider a pulse of rectangular
profile, i.e., $R(t)=R$ at $0<t<T$ with pulse duration as long as
$T$, and take the strict resonance condition when the excitation
frequency is exactly the transition frequency providing
$\Delta_{21}=0$.

The solution for this nonlinear set of equations \rf{Res3} we will
find using the method of successive approximations and be accurate
to the first order approximation only. In case of exact resonance
the zero approximation (no local field) for the equation has the
form:
 \be
  \dsp{i\frac{d\ra^{(0)}}{dt}}-[\Va,\ra^{(0)}]=0, \\
 \ee
and its solution is found easily:
 \be
   \label{Zerro}
   \ra^{(0)}(t)=\frac{1}{2}\left(
   \begin{array}{cc}
     1-cos(2Rt) & sin(2Rt) \\
     -sin(2Rt)  & 1+cos(2Rt) \\
   \end{array}
   \right).
 \ee
In the first order approximation the equation is
 \be
  \dsp{i\frac{d\ra^{(1)}}{dt}}-[\Va+\Vp(\ra^{(0)})]=0, \\
 \ee
which is as well integrated:
 \be
   \label{Zerro1}
   \ra^{(1)}(t)=\frac{1}{2}\left(
   \begin{array}{cc}
     1-cos(V(t)) & sin(V(t)) \\
     -sin((V(t))  & 1+cos((V(t)) \\
   \end{array}
   \right),
 \ee
 $$
  \dsp{V(t)=2Rt-B(1-cos(2Rt))},
   $$
where $\dsp{B=\frac{3n(\rk)G\lambda_{21}^3}{8\pi^2}
 \frac{\gamma}{4R}}$. Substituting results of \rf{Zerro1} in \rf{Res3}
for $\beta^{'}(t)$ and neglecting the signal retardation we get
 \be
   \label{Rad}
   \begin{array}{c}
    \dsp{\beta^{'}(t)=M e^{-i\varepsilon_\bk U}\Biggl[\int_0^{U_e}e^{i\varepsilon_\bk
    u}sin(u+B(1-cos(u)))du+  }\\
    \dsp{ \int_{U_e}^{U}e^{i\varepsilon_\bk u}sin(U_e+B(1-cos(U_e))) du\Biggr] }\\
    \dsp{ 2R\tau=u,\;\;2Rt=U,\;\;
    \;\;U_e=2RT,} \\
    \dsp{ M=\frac{N}{2R}\sqrt{\frac{2\pi\omega_\bk}{\hbar W}} ||{\mu}_{21}||,\;\;\;\;\frac{\nu_\bk-i\eta_\bk/2 }{2R}=\varepsilon_\bk },\\
   \end{array}
 \ee
where $N$ is the number density in the sample. The second part in
the expression \rf{Rad} describes radiation of residual
polarization induced during pulse propagation. It has the nature
of ordinary Rayleigh scattering. It is of no particular interest
for the present study so we will disregard its existence in the
following analysis. Now, using the following relation
\cite{Ryzik}:
  \be
    \label{Rad1}
    \begin{array}{c}
      \dsp{cos(B cos(u))=\sum_{j=-\infty}^{\infty}(-1)^j J_{|2j|}(B)e^{i2ju}}, \\
      \dsp{sin(B cos(u))=\sum_{j=-\infty}^{\infty}(-1)^j
      J_{|2j+1|}(B)e^{i(2j+1)u}},
    \end{array}
  \ee
the integral \rf{Rad} can be presented as
  \be
    \label{Rad2}
    \begin{array}{c}
     \dsp{\beta^{'}(t)=M e^{-i\varepsilon_\bk U}\int_0^{U_e}e^{i\varepsilon_\bk
     u}\times }
      \dsp{ \sum_{j=-\infty}^{\infty}(-1)^j
      e^{i ju} Y_j} , \\
       Y_{2j+1}=\bigl(J_{|2j|}e^{iB}+J_{|2j+2|}(B)e^{-iB}\Bigr),\\
      Y_{2j}=\bigl(J_{|2j|+1}(B)e^{iB}-J_{|2j|+3}(B)e^{-iB}\bigr),\\
      Y_{-2j}=Y_{2j}^{*},\;\;\ Y_0= 2J_{1}(B)cos(B).

    \end{array}
  \ee
Performing integration we get
  \be
    \label{Rad3}
    \begin{array}{c}
     \dsp{\beta^{'}(t)=M e^{-i\varepsilon_\bk U}
      \sum_{j=-\infty}^{\infty}(-1)^j S(j)Y_j,} \\
      \dsp{S(m)=\frac{e^{i(\varepsilon_\bk+m)U_e}-1}{i(\varepsilon_\bk+m)}}.
     \end{array}
  \ee
Substituting \rf{Rad3} and \rf{Rad} into \rf{Res5} and dropping
all non resonant cross terms we can write the spectrum of
scattered light in the form:
  \be
    \label{Rad4}
    \begin{array}{c}
      \dsp{ {I}_\bk(t)=\eta_\bk\hbar\omega_\bk
      |\beta_\bk^{'}(t)|^2
      =}
       \dsp{  N ^2 G_\bk^2 e^{-\eta_\bk t_1}
     \sum_{j=-\infty}^{\infty}|S(j)|^2|Y_j|^2 } \\
   \dsp{|S(m)|^2=\frac{1+e^{-\eta_\bk T}-2e^{-\eta_\bk
   T/2}cos((\nu_\bk+2mR)T)}{(\nu_\bk+2mR)^2+\eta_\bk^2/4},}\\
   \dsp{G_\bk={\eta_\bk\hbar\omega_\bk\frac{2\pi\omega_\bk}{\hbar W}
   ||{\mu}_{21}||^2}},  \\
    \end{array}
  \ee
where $t<T \to t_1=0,\;\;\; t>T \to t_1=t-T$.
Let us now see this expression in various marginal cases.

  1) $\eta_\bk T \gg 1$. In this limit the contribution from the terms
containing $e^{-\eta_\bk T}$ is negligible over long times, so
\rf{Rad3} takes the form:
  \be
    \label{Rad5}
    \begin{array}{c}
      \dsp{{I}_\bk(t)= N^2 G_\bk^2 e^{-\eta_\bk t_1} }
      \dsp{\Biggl[
      \sum_{j=-\infty}^{\infty}\frac{|Y_j|^2}{(\nu_\bk+2jR)^2+\eta_\bk^2/4}
   \Biggr].} \\
    \end{array}
  \ee
The spectrum represents a collection of lines multiple to Rabi
frequency with line widths of the order of $\eta_\bk^2/4$ and
intensities proportional to $|Y_j|^2$. Line widths and intensities
as functions of $B$ are shown in \rff{fig:1} for several central
lines. It must be noted that such analysis is reasonable if
carried out at the value of parameter $B \ll 1$ as the result of
this section were obtained using the method of successive
approximations in $B$. As it follows from the picture the line
intensities decrease sharply with larger values of $j$ which, in
its turn, follows from the asymptotic expansion for Bessel
functions:
  $
   \dsp{J_j(B)\sim\frac{1}{\sqrt{2\pi j}} \Biggl(\frac{e
   B}{2j}\Biggr)^j}
  $
\begin{figure}
    \includegraphics[scale=1.5]{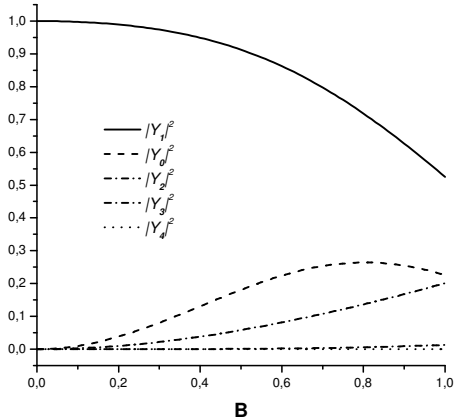}
   \caption{Relative intensities of spectral lines.}
  \label{fig:1}
  \end{figure}
2) $\eta_\bk T \ll 1$. In this limit it is valid to put
$e^{-\eta_\bk T}\sim 1$ in the expression for the intensity
spectra to change it to
    \be
    \label{Rad6}
    \begin{array}{c}
      \dsp{{I}_\bk(t)=
       N^2 G_\bk^2 e^{-\eta_\bk t_1} }
      \dsp{\sum_{j=-\infty}^{\infty}2|Y_j|^2 V(j) } , \\
   \dsp{V(m)=\frac{1-cos((\nu_\bk+2mR)T)}{(\nu_\bk+2mR)^2+\eta_\bk^2/4}}.
    \end{array}
  \ee
Over long times the spectrum of scattered light is now as well a
collection of lines multiple to Rabi frequency with intensities
proportional to $|Y_j|^2$. However, unlike the opposite case they
are modulated with the frequency inverse to the pulse duration
time.  This modulation the line widths are of the order of inverse
pulse duration time. For the central peak it is seen from the
following expression \footnote{similarly for other components}:
  \be
    \frac{1-cos(\nu_\bk T)}{\nu_\bk^2+\eta_\bk^2/4}=\frac{\sum_{i=1}^{\infty}(-1)^{i-1}(\nu_\bk T)^{2i}/(2i)!}{\nu_\bk^2+\eta_\bk^2/4}
  \ee
for large values of $\nu_\bk\gg\eta_\bk$ the expression takes the form:
  \be
    \label{Modul}
   \frac{1-cos(\nu_\bk T)}{\nu_\bk^2+\eta_\bk^2/4}\approx
   T^2/2!-\nu_\bk^2T^4/4!+ O(\nu_\bk^4),
  \ee
from which it follows that the line width value is of the order of $1/T$.
Let us note that the intensity in the line center is always different from null. Indeed, at $\nu_\bk \to 0$ we have:
  \be
     \frac{1+e^{-\eta_\bk T}-2e^{-\eta_\bk T/2}cos(\nu_\bk T)}{\nu_\bk^2+\eta_\bk^2/4} =
     T^2+O(T^4)
  \ee

\section{Numerical modeling \label{Sec:4}}
In the previous section we obtained explicit expressions for
intensity spectra of scattered radiation in case when parameter
$B\ll 1$ which corresponds to the small density of atoms. This
limitation was due to the method of successive approximations
selected for our analysis. In case of large density values this
method is not applicable. Thus, we have to solve the system of
nonlinear system equations \rf{Res3} using numerical methods.

In order to estimate the range of parameters where multicomponent
spectra could be expected we introduced operators describing
radiative and collisional relaxation.
 \be
   \label{Num1}
    \begin{array}{c}
 \dsp{     \Gamma[\ra]=-i\gamma/2(\ptt \ra + \ra \ptt - 2\pot\ra\pto),
     } \\
    \dsp{ \Gamma_s[\ra]=-i\gamma_2(\pot \rai{12}+\pto \rai{21})}.
    \\
     \end{array}
 \ee
Substituting these damping terms into the equations gives
 \be
  \label{Res6}
  \begin{array}{c}
    \dsp{i\frac{d\ra}{dt}}-[\Ha+\Va+\Vp,\ra]-\Gamma[\ra]-\Gamma_s[\ra]=0, \\
   \dsp{  \frac{d\beta_\bk^{'}}{dt}=-(i\nu_\bk+\eta_\bk/2)\beta_\bk^{'}-\sum_m^N\gkm^{*}\rai{21}},\\
    \dsp{{I}_\bk(t)=\eta_\bk\hbar\omega_\bk |\beta_\bk^{'}(t)|^2. }
  \end{array}
 \ee
Also, we present the time integrated intensity spectra to make
experimental verification of obtained results possible if such is
to take place.
 \be
   I_\bk^T=\int_0^\infty{I}_\bk(t)dt.
 \ee
In ris. 2 one can see the intensity spectra of scattered light for
a long compared to the mode dumping time pulse \footnote{This case
is examined as case 1) in the Sec. \ref{Sec:3}} , $\eta_\bk T \gg
1$, for different values of parameter $B$. For small values of $B$
in ris. 2(a) ot is seen that the ratios of line intensities
generally meet the results reported in the previous section.  The
central peak with $\nu_\bk=0$ is the only exception. This fact is
explained by existence of residual radiation after the pulse has
already passed the medium. With increasing value of $B$ the
relative intensity of the central line grows substantially. The
relative intensities of the side bands decrease while shifting to
the center of the spectral range, see \rff{fig:2}(b),(c).
Moreover, one can see that peaks multiple to 3 and 4 Rabi
frequencies arise shifted significantly to the center. At
extremely large densities $B\sim1$ all components merge leaving
the only central peak (\rff{fig:2}(d)).

\begin{figure*}
    \includegraphics[scale=1.8]{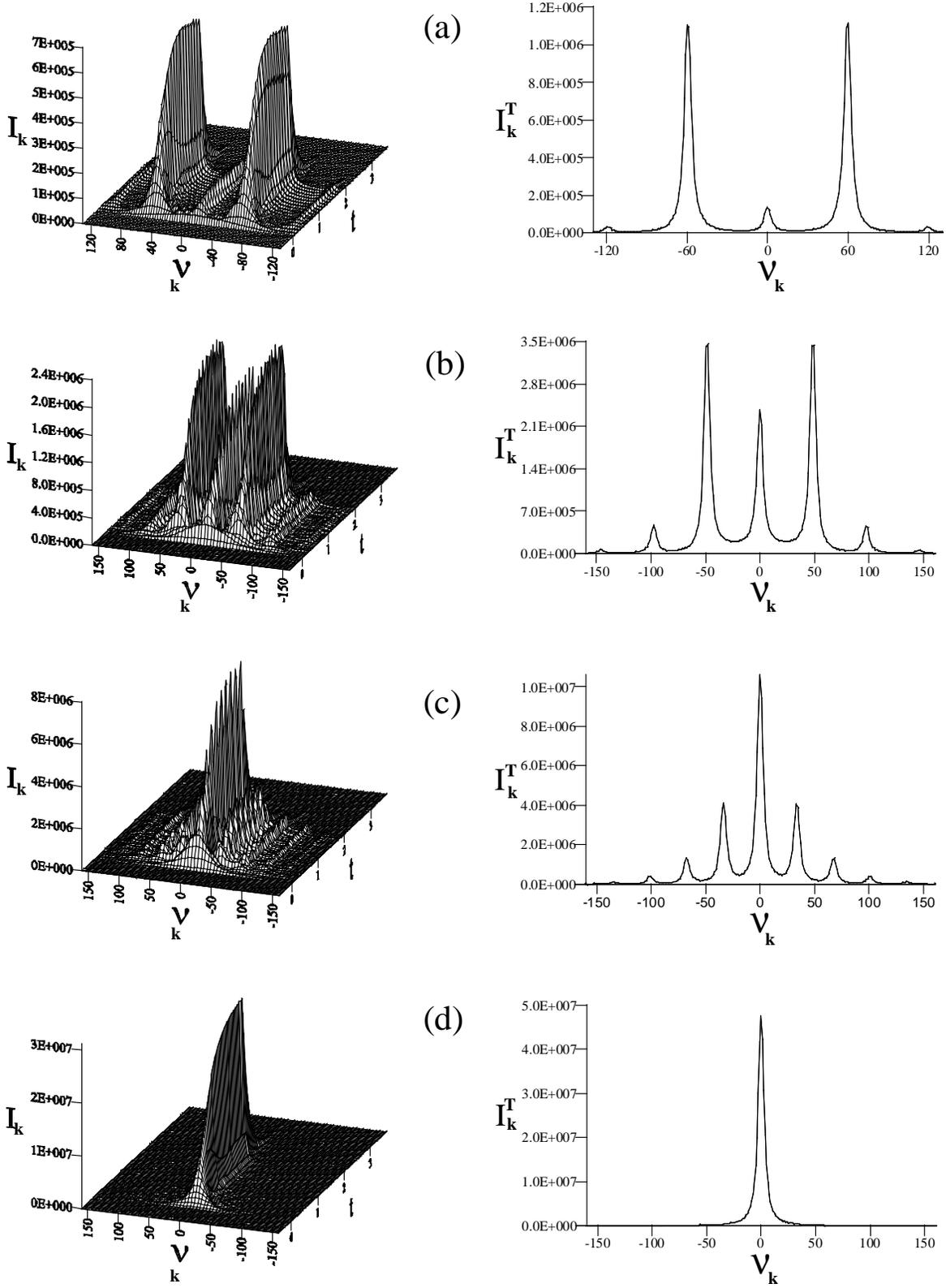}
   \caption{ Intensity spectrum of the scattered light $I_\bk$ and time integrated spectrum
   $I_\bk^T$, values of $I_\bk,I_\bk^T$ are presented in relative units and proportional to
   $N^2$. $\Delta=0$,$R=10\pi$,$\gamma=0.01$, $\eta_\bk=5$, $T=1.9$
   $B=0.32,\;0.64,\;0.85,\;1.2$ for pictures (a,b,c,d) respectively.}
  \label{fig:2}
  \end{figure*}

The short pulse scattering, $\eta_\bk T \ll 1$, is demonstrated in
\rff{fig:3} for different values of density. In \rff{fig:3}(a),(b)
some additional peaks are well seen. These peaks arise due to
modulations of the kind \rf{Modul} for $6\pi$ and $10\pi$ pulses
respectively. The general behavior of spectrum is just like the
one displayed in the previous case, that is, the peaks shift
toward the central line at $B\sim 1$ and finally merge.
 \begin{figure*}
    \includegraphics[scale=1.8]{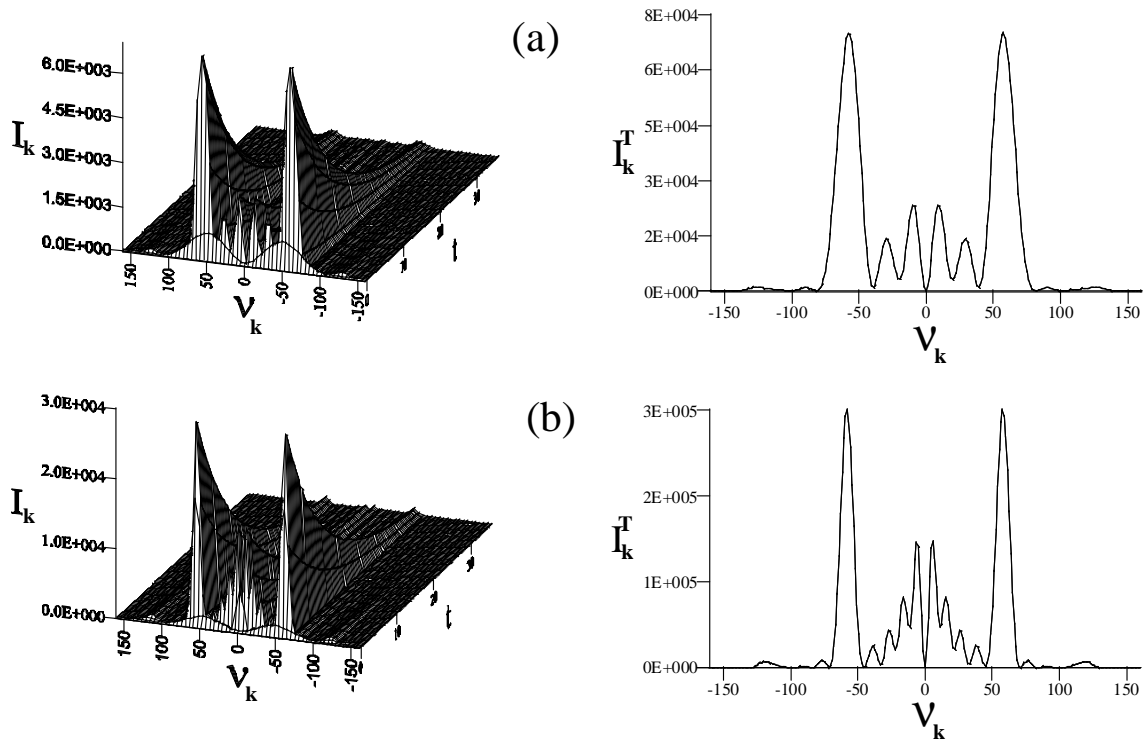}
   \caption{Same as in \rff{fig:2}. $\Delta=0$,$R=10\pi$,$\gamma=0.01$, $\eta_\bk=0.1$,
   picture (a) $T=0.3$, $B=0.28$, picture (b) $T=0.5$, $B=0.36$}
  \label{fig:3}
  \end{figure*}
The calculation carried out for different values of
$\gamma,\gamma_2$showed that with shorter relaxation time the
spectral picture is "blurred". At $\gamma,\gamma_2 \sim 2/T$ the
additional peaks induced by modulation practically vanish.
However, the peaks multiple to $\pm 4R$ remain observable up until
$\gamma_2 T \sim 8$ (\rff{fig:4}).
\begin{figure*}
    \includegraphics[scale=1.8]{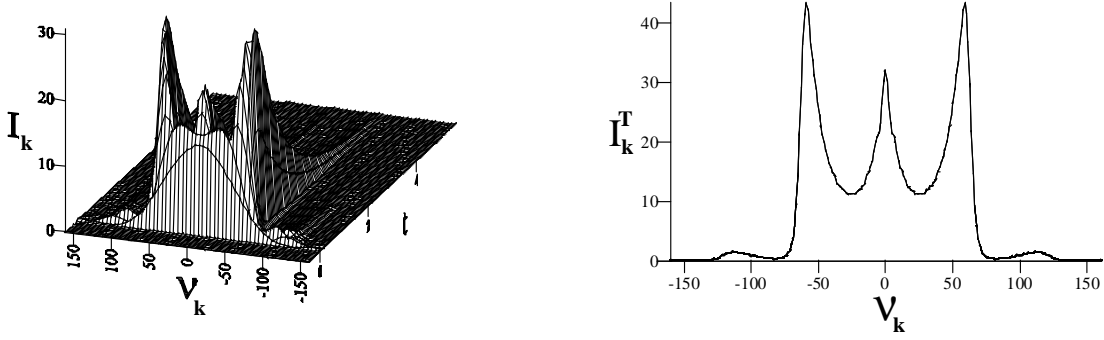}
   \caption{Same as in \rff{fig:2}. $\Delta=0$,$R=10\pi$,$\gamma=0.5$,$\gamma_2=9$,$\eta_\bk=1$,
   $T=1$, $B=0.74$}
  \label{fig:4}
  \end{figure*}
We examined the spectra as functions of detuning and found that
the spectral pattern is very sensitive to the value of detuning.
Even for $\Delta\sim ~0.1R$ the spectrum is distorted showing
increased intensities of the lines lying close to the excitation
frequency. At $\Delta\sim ~0.8R$ the multimode structure is no
longer observed. The light is scattered at the frequency of the
pulse.

\section{Summary \label{Sec:5}}
In this work we have carried out the analysis of scattering of a
short laser pulse in a dense two-level medium. We used the
approach based on the BBGKY-hierarchy for the reduced density
operators. In the framework of the Hartree-Fock approximation and
with use of the properties of coherent state operators we derived
the cooperative field operator in the explicit form and compared
our results with the well-known ones \cite{Bowden_79}. The
transformations used in this work allowed us to derive the set of
equations describing transient spectra of scattered light. The
equations were analyzed using the method of successive
approximations in the case when the medium was excited by a short
rectangular laser pulse. A possibility was demonstrated to
generate scattered light at frequencies multiple to Rabi frequency
which is contributed by transient local field. The ratio of line
intensities has been determined. We also demonstrated that
additional satellites can occur in the spectrum due to modulation
Rabi components as a result of finite pulse length.

In the final part of this work we carried out the numerical
analysis of spectra. It was demonstrated that in the limit of
small density the theoretical analysis well meets the results of
calculations. However, for large densities, i.e., $B>0.3$ this
theoretical approach becomes inapplicable. As the density is
increased one can observe a significant shift of the satellites to
the central line in came cases accompanied with generation of
additional bands multiple to three and four Rabi frequencies.

Authors are grateful to A.N. Starostin for discussions and remarks
received in the course of work. We would like to acknowledge the
financial support from the Russian Foundation for Basic Research
No.02-02-17153 and No.03-02-06590 and grants of the President of
the Russian Federation No.MK1565.2003.02, No.MD338.2003.02 and
No.NS1257.2003.02.

  \setcounter{equation}{0}

\section*{Appendix A. Cooperative field operator}
\renewcommand{\theequation}{A.\arabic{equation}}

In this section we will derive the explicit form for the cooperative field operator.
Our attention is now for the expression~\rf{Zamena1}:
 \be
  \label{koop1}
  \begin{array}{c}
   \dsp{[\tilde{V}_{af}^{p},\rft \rho_{a}]\Spf=\sum_{\bkl}  i} \dsp{ [\gk\pto\beta_\bk(t)^{'}}-\\-\dsp{ \gk^{*}\pot\beta_\bk^{'*}(t), \rft \ra], }\\
 \end{array}
 \ee
where the equation for $\beta_\bk(t)$ has the form:
 \be
   \dsp{  \frac{d\beta_\bk^{'}}{dt}=-(i\nu_\bk+\eta_\bk/2)\beta_\bk^{'}-\sum_m^N\gkm^{*}\rai{21}}.\\
 \ee
Solving the equation for $\beta_\bk(t)^{'}$ formally one can get:
 \be
  \dsp{\beta_\bk^{'}(t)=-
  \sum_l^N\gkm^{*}\int_{0}^{t}\rai{21}(\tau)e^{-i(\nu_\bk+\eta_\bk/2)(t-\tau)
  }d\tau}.
 \ee
Now we substitute the result in \rf{koop1} and by changing the
summation over atomic and field variables for integration
correspondingly over the volume and $\bk$ come to
 \be
   \label{koop2}
   \begin{array}{c}
    \dsp{[\tilde{V}_{af}^{p},\rft \rho_{a}]\Spf=[R_{l}(t,\rk_l)\pto - R_{l}^{*}(t,\rk_l)\pot, \rft \ra] }\\
   \dsp{R_{l}(t,\rk_l)=\int \int_{0}^{t}B_{lm}(kr_{lm}) \times} \\
    \dsp{\times\rai{21}(\tau,\rk_m)e^{-i(\nu_\bk+\eta_\bk/2)(t-\tau)}d\tau}d\rk_m k^2dk,\\
   \end{array}
 \ee
 \be
   \label{koop3}
   \begin{array}{c}
   \dsp{B_{lm}(kr_{lm})= \int\sum_{\alpha=\pm 1} \gk\gkm^{*}d\Omega, }
    \end{array}
 \ee
where $d\Omega$ is the spatial angle element, $k=|\bk|$,
$r_{lm}=|\rk_l-\rk_m|$.  Integral of the kind \rf{koop3} have been
considered in \cite{Knight} and are representable in the following
form:
  \be
   \label{koop4}
   \begin{array}{c}
   \dsp{ B_{ml}(x)=\frac{2\pi\omega_\bk||{\mu}_{21}||^2 }{\hbar W}}\times \\
   \times\dsp{  \Biggl(  b\frac{sin (x)}{x}+c\Bigl(\frac{sin (x)}{(x)^3}-\frac{
   cos(x)}{(x)^2}\Bigr)\Biggr)}.\\
    \end{array}
 \ee
Transition constants $b,c$ for various polarizations
$\Delta M$ are
  \be
   \label{koop5}
   \begin{array}{c}
   \dsp{\Delta M=0: \;\;\;\;\;b=0,c=2,} \\
    \dsp{\Delta M=\pm1: \;\;\;\;\; b=1,c=-1.} \\
    \end{array}
 \ee
With use of \rf{koop4} the integral \rf{koop2} was examined in
\cite{Bowden}, where allowing for disregard of signal retardation
they obtained the cooperative field operator:
 \be
  \Vp=\dsp{-i\frac{3n(\rk)G\lambda_{21}^3}{8\pi^2} \gamma/2\;\pto
  \rai{21}(t,\rk_l)+ H.c., }
 \ee
where $\gamma=\dsp{\frac{4\omega_{21}^3||\mu_{21}||^2}{3\hbar
c^3}}$ is the spontaneous decay rate from the upper state,
$n(\rk)$ is the density of the sample, and $G$ represents the
geometrical factor being a function of selected geometry of the
sample, its dimensions, and polarization of scattered light.

\end{document}